\documentclass[epj,final]{svjour}
\usepackage{graphicx,amstext}
\newcommand{\vs}{{\it vs.\ }}
\newcommand{\al}{{\it et al.}~\cite}
\newcommand{\pff}{(TMTSF)$_2$PF$_6$}
\begin{document}
\sloppy
\title{Influence of Quantum Hall Effect on Linear and Nonlinear Conductivity in the FISDW States of the Organic Conductor (TMTSF)$_2$PF$_6$}
\titlerunning{Quantum Hall effect breakdown and FISDW sliding}
\author{T.~Vuleti\'{c}\inst{1,2} \and C.~Pasquier\inst{1} \and P.~Auban-Senzier\inst{1} \and S.~Tomi\'{c}\inst{2} \and D.~J\'{e}rome\inst{1} \and K.~Maki\inst{3} \and K.~Bechgaard\inst{4} \mail{tvuletic@ifs.hr}}
%
\institute{Laboratoire de Physique des Solides, Universit\'{e} Paris-Sud, F-91405 Orsay, France \and Institute of Physics, P.O. Box 304, HR-10000 Zagreb, Croatia \and Dept.~of Physics \& Astronomy, University of Southern California, Los Angeles, CA 90089-0484 \and Dept.~of Condensed Matter Physics and Chemistry, Ris\o~National Laboratory, DK-4000 Roskilde, Denmark}
\date{Received: date / Revised version: date}
\abstract{
We report a detailed characterization of quantum Hall effect (QHE) influence on the linear and non-linear resistivity tensor in FISDW phases of the organic conductor (TMTSF)$_2$PF$_6$. We show that the behavior at low electric fields, observed for nominally pure single-crystals with different values of the resistivity ratio, is fully consistent with a theoretical model, which takes QHE nature of FISDW and residual quasi-particle density associated with different crystal imperfection levels into account. The non-linearity in longitudinal and diagonal resistivity tensor components observed  at  large electric fields reconciles preceding contradictory results. Our theoretical model offers a qualitatively good explanation of the observed features if a sliding of the density wave with the concomitant destruction of QHE, switched on above a finite electric field, is taken into account. 
}  
\maketitle
\section{Introduction} 
Highly anisotropic organic superconductor  tetramethyltetraselenafulvalene -- hexafluorophosphate, \pff~exhibits interesting properties in all  regions of  the phase diagram \cite{cha12}. All yet discovered electronic transport mechanisms are found in the low temperatures, high pressure regime (below 12~K, up to and above 8~kbar \cite{cha12,Ishiguro,kwak45,bisk15}). Besides metallic conductivity and superconductivity (SC) the other two, theoretically dissipationless, electrical transport  mechanisms occur in the spin-density wave (SDW) state and  the quantum Hall effect (QHE) state, respectively.  An SDW  phase is established below 12~K at ambient pressure. An applied pressure decreases the SDW transition temperature and finally at pressures higher than 8~kbar superconductivity appears below about 1.2~K.  A small magnetic field destroys superconductivity and  above a threshold  magnetic field a cascade of field induced spin density wave phases (FISDW) appears. 
FISDW phases differ from the ambient-pressure SDW phase in that they are semimetallic with QHE occuring in each phase. The electric field dependent transport has been studied by several authors and disparate results have been obtained (\cite{Osada56,Kang34,Balicas08}). 
Osada {\it et al.}~\cite {Osada56} were the first to investigate the subject of nonlinear conduction in FISDW phases. They studied (TMTSF)$_2$ClO$_4$ and reported that the longitudinal resistivity $\rho_{xx}$ increases above a negligibly small threshold electric field, whereas the Hall resistivity $\rho_{xy}$ remains constant with electric field. They used the six contact geometry, in which two current contacts covered the ends of the sample, whereas four voltage contacts were mounted in the way that two were attached on each $a$-$c^*$ face of the sample. Later, Kang {\it et al.}~\cite {Kang34} found the opposite result for (TMTSF)$_2$PF$_6$, that is a longitudinal resistivity $\rho_{xx}$ decrease above a finite electric field. Since the latter authors used the four annular contacts geometry, they were not able to measure $\rho_{xy}$. Finally, Balicas~\cite{Balicas08} reported a strong increase of the longitudinal resistivity $\rho_{xx}$  and a steep decrease of the  Hall resistivity $\rho_{xy}$  above a finite electric field for the \pff~material. He used the eight contact geometry, with a pair of current and a pair of voltage contacts on each $a$-$c^*$ face of the sample.

In analogy with the ambient pressure SDW phase of \pff~it is expected for the FISDW phases to exhibit nonlinear conductivity of the same origin. That is, nonlinear conductivity due to the sliding of the SDW condensate depinned above the threshold electric field $E_T$. This issue has been theoretically described by  Virosztek and Maki (VM)~\cite{ViMa75}.  An experimental observation of Kang {\it et al.}~\cite {Kang34} was in accordance with the VM theoretical prediction. However, Virosztek and Maki completely disregarded the  QHE nature of the FISDW phases,  hallmarked by the nondiagonal, Hall  component of the conductivity tensor, which is quantized and becomes  large, $\sigma_{xy}^Q=-2Ne^2/h$. In addition, although the longitudinal conductivity $\sigma_{xx}$ goes to zero as in a gapped semiconductor, the longitudinal resistivity $\rho_{xx}$, also goes to zero, due to its tensorial nature which  is dominated by a large value of $\sigma_{xy}^Q$ at low temperatures.  $\rho_{xx}$ goes to zero at low temperatures because  the denominator is dominated by ${\sigma_{xy}^Q}^2$, which is finite at low temperatures, whereas $\sigma_{yy}\rightarrow0$ (see later Eqs.~\ref{rxx},~\ref{rxy},~\ref{ryy}). The same applies to $\rho_{yy}$. The QHE nature of FISDW phases  was recently taken into account by Yakovenko and Goan~\cite{Yako81}. They have shown that the FISDW motion under the applied electric field results in a destruction of the QHE state. Indeed, a concomitant increase and decrease of the longitudinal resistivity and the Hall resistivity, respectively, observed by Balicas~\cite{Balicas08} qualitatively confirm the validity of this approach. Conversely, an electric-field independent Hall resistivity concomitant with the rise of longitudinal resistivity found by Osada {\it et al.}~\cite {Osada56} cannot be understood in the framework of the existing theoretical models \cite{ViMa75,Yako81,Yako82}. 

The aim of present work is to provide a unified picture of electric field-dependent transport in FISDW phases. Our results establish a missing link between the already observed opposite behaviors of the longitudinal resistivity as a function of electric  field. We show that experiments of  Kang {\it et al.} and Balicas have been designed to investigate a limited range of the 4-dimensional phase space (pressure-temperature-magnetic field-electric field) in which dominates either QHE or classical SDW nature of FISDW phases, respectively. Further, our results reveal decisive influence of the sample quality and the contact configuration used in the experiment on the observed behavior of the  resistivity tensor.  We show that the experimentally observed behavior at low electric fields of the longitudinal component  $\rho_{xx}$, and the Hall component  $\rho_{xy}$, of the resistivity tensor  in each FISDW subphase, is perfectly consistent with the theoretical prediction, which takes  QHE nature of FISDW into account. Finally, we argue that the latter feature incorporated in a theoretical model for sliding FISDW offers a qualitatively new and very plausible interpretation of the observed non-linear behaviour of longitudinal resistivity.

\section{Experimental and Results}
 \label{exp}
Nominally pure single crystals of \pff, originating from different batches had standard sample dimensions $3\times 0.2 \times 0.1 \textrm{mm}^3$, which were imposed by the pressure cell diameter.  Samples were visually aligned with  the magnetic field $H$, so that it was parallel to $c^*$ within $5^o$.   Linear regime measurements were performed by the standard AC lock-in technique. Nonlinear curves at fixed points in $(T,H)$ phase diagram were taken by a pulse technique in order to avoid heating of the sample due to very high currents used. Maximum applied currents were 65mA.\@ Various pulse lenghts were used: 10, 20 or 40 $\mu$s with a repetition period of 40 ms, which was at least thousand times longer than the pulse. Non-heating was checked by the shape of the pulse displayed on the osciloscope, and by measurements at $(T,H)$ points outside FISDW phase, where the linear conductivity was regularly obtained as expected.  The antisymmetrical average of $R_{xy}$ with respect to both magnetic field directions was taken to suppress the superimposed magnetoresistance signal. Accordingly, the symmetrical average of $R_{xx}$ and $R_{yy}$ was always taken to suppress superimposed Hall signal.

We have studied four single crystals with different resistivity ratios $RR=\rho_{RT}/\rho_{min}$, where $\rho_{RT}$ and $\rho_{min}$ are resistivities measured at room temperature and at 1.5K, just above SC transition, respectively. Two single crystals with $RR\sim40$ and two with $RR\sim1000$ have been labelled as low quality and high quality samples, respectively. Two distinct contact configurations have been used. In the first configuration, four annular contacts have been attached to the sample and the longitudinal resistivity $\rho_{xx}$ was measured. In the second, eight contact configuration, four contacts were attached on each of $a$-$c^*$ faces of the single crystal. The latter contact arrangement allowed the measurement of  $\rho_{xx}$, $\rho_{xy}$ and $\rho_{yy}$ components of the resistivity tensor.

Our first important result concerns the influence of sample quality on the  components of resistivity tensor in the FISDW phases. First, we present the behaviour obtained in the linear regime as a function of the sample quality, and then we show the corresponding behaviour in the non-linear regime.

\begin{figure}
\centering\resizebox{0.42\textwidth}{!}{\includegraphics*{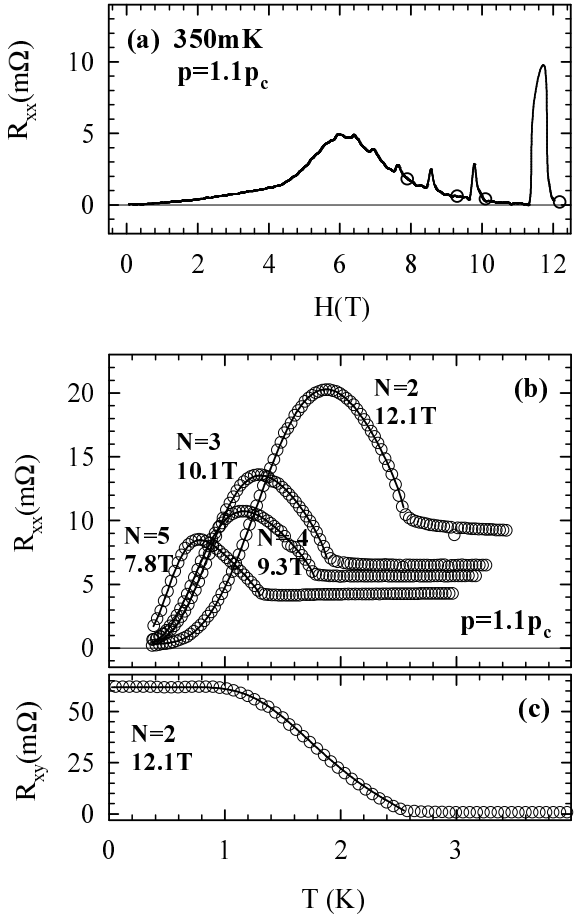}}
\caption{High quality sample: (a) Longitudinal resistance  $R_{xx}$ \vs magnetic field $H$ at the lowest temperature.  Circles define magnetic fields at which  temperature sweeps presented in (b) were performed. (b) Longitudinal resistance  $R_{xx}$ \vs temperature $T$ at fixed  magnetic fields. (c) Hall resistance, $R_{xy}$ \vs temperature $T$ at field of 12.1 T.  Open points in (b) and (c) represent experimental data and solid lines represent the  fits based on the theory (see text).} 
\label{high}
\end{figure}

In Fig.~\ref{high} we show the longitudinal resistance $R_{xx}$ and the Hall resistance $R_{xy}$ as a function of magnetic field and temperature observed for a high quality single-crystal. The observed curves reproduce the well known behaviour established in FISDW phases by different authors \cite{cha12,Balicas08,KangCha,coop22}. That is, dips in the resistance occuring inside the FISDW phases and peaks between the phases are pronounced, Fig.~\ref{high}(a). At the lowest temperatures (see Fig.~\ref{high}(b)), the residual longitudinal resistance inside FISDW phase approaches zero value, whereas the Hall resistance attains the enhanced, precisely defined "plateau" value. 

\begin{figure}
\centering\resizebox{0.42\textwidth}{!}{\includegraphics*{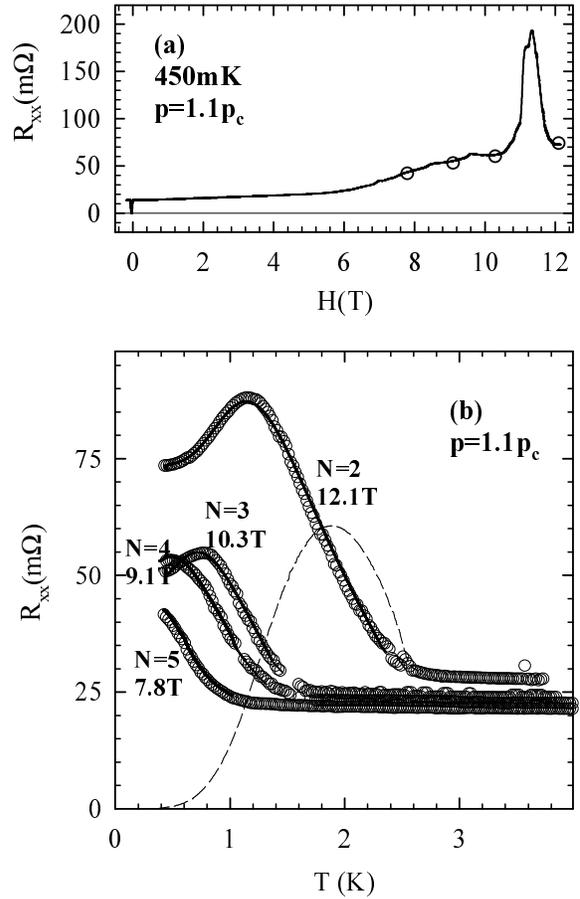}}
\caption{Low quality sample: (a) Longitudinal resistance  $R_{xx}$ \vs magnetic field $H$ at the lowest temperature.  Circles define magnetic fields at which  temperature sweeps presented in (b) were performed. (b) Longitudinal resistance  $R_{xx}$ \vs temperature $T$ at fixed  magnetic fields.  An $R_{xx}$ curve at 12.1 T for a high quality sample, normalized to the resistance of low quality one in the normal state,  is shown by dashed line for a comparison. Open points in (b) represent experimental data and solid lines represent the  fits based on the theory (see text).}
\label{low}
\end{figure}

Results for a low quality single crystal are presented in Fig.~\ref{low}. First feature which reflects low sample quality is that dips and peaks are smeared out in the curve of longitudinal resistance $R_{xx}$ \vs magnetic field $H$ (Fig.~\ref{low}(a)).  Second feature is visible in $R_{xx}$ \vs $T$ curves at fixed magnetic fields,  shown in Fig.~\ref{low}(b). That is, the longitudinal resistance measured at the lowest temperatures is higher  than the resistance at 3.2 K in the normal phase. However, as far as $T_{c}(N)$, the transition temperature to the corresponding FISDW phase is concerned, we observed similar values for respective phases in low, as well as in high quality samples (Fig.~\ref{high}(b), (c) and Fig.~\ref{low}(b)).

\begin{figure}
\centering\resizebox{0.42\textwidth}{!}{\includegraphics*{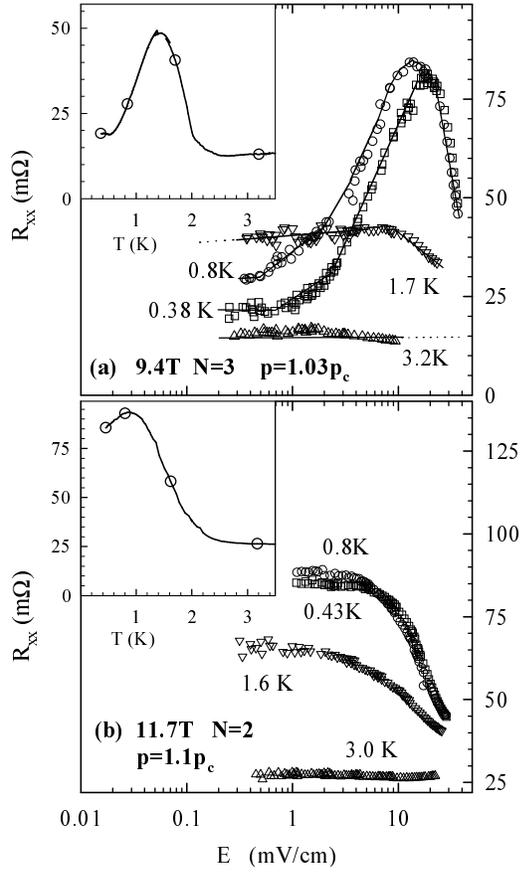}}
\caption{Longitudinal resistance $R_{xx}$ \vs electric field $E$  at different temperatures  for (a) a high quality sample and for (b) a low quality sample. Solid lines are guides for the eye. Inset: Linear regime temperature sweeps at corresponding magnetic fields. Circles denote temperatures at which  nonlinear measurements were performed.} 
\label{NonL9900}
\end{figure}

Further, we show the electric-field dependent longitudinal resistance $R_{xx}$ for hereabove described single crystals of different quality. 
A high quality sample showed the following behavior (Fig.~\ref{NonL9900}(a)). At 0.38 K and 0.8 K, {\it i.e.} at temperatures below the peak in the $R$ \vs $T$ low field curves (see Inset), we observed a  substantial rise of the resistance above a finite threshold electric field of 1 and 0.5 mV/cm, respectively.   $R_{xx}$ displays a maximum at fields of about 20 and 15 mV/cm, and falls down rapidly. At 1.7 K, {\it i.e.}  at temperatures above the peak in the $R$ \vs $T$ low field curve, $R_{xx}$ showed qualitatively the same, but much less pronounced behaviour. That is, only a very weak increase switched on by a very low field not higher than 0.5 mV/cm, and a crossover to a final decrease at field of 8 mV/cm. Finally, at 3.2 K in the normal phase the longitudinal resistance stays constant up to 10 mV/cm, which is about an order of magnitude larger field than the highest threshold detected in the FISDW phase. 

For a low quality sample the field dependence of the longitudinal resistance is constant until a threshold field is reached, above which it decreases (Fig.~\ref{NonL9900}(b)). The threshold field is larger and the electric field dependence is smoother than for a high quality one. In addition, $R_{xx}$ stayed constant in the normal phase up to 25 mV/cm, which is about an order of magnitude larger field than the highest threshold detected in the FISDW phase.

\begin{figure}
\centering\resizebox{0.42\textwidth}{!}{\includegraphics*{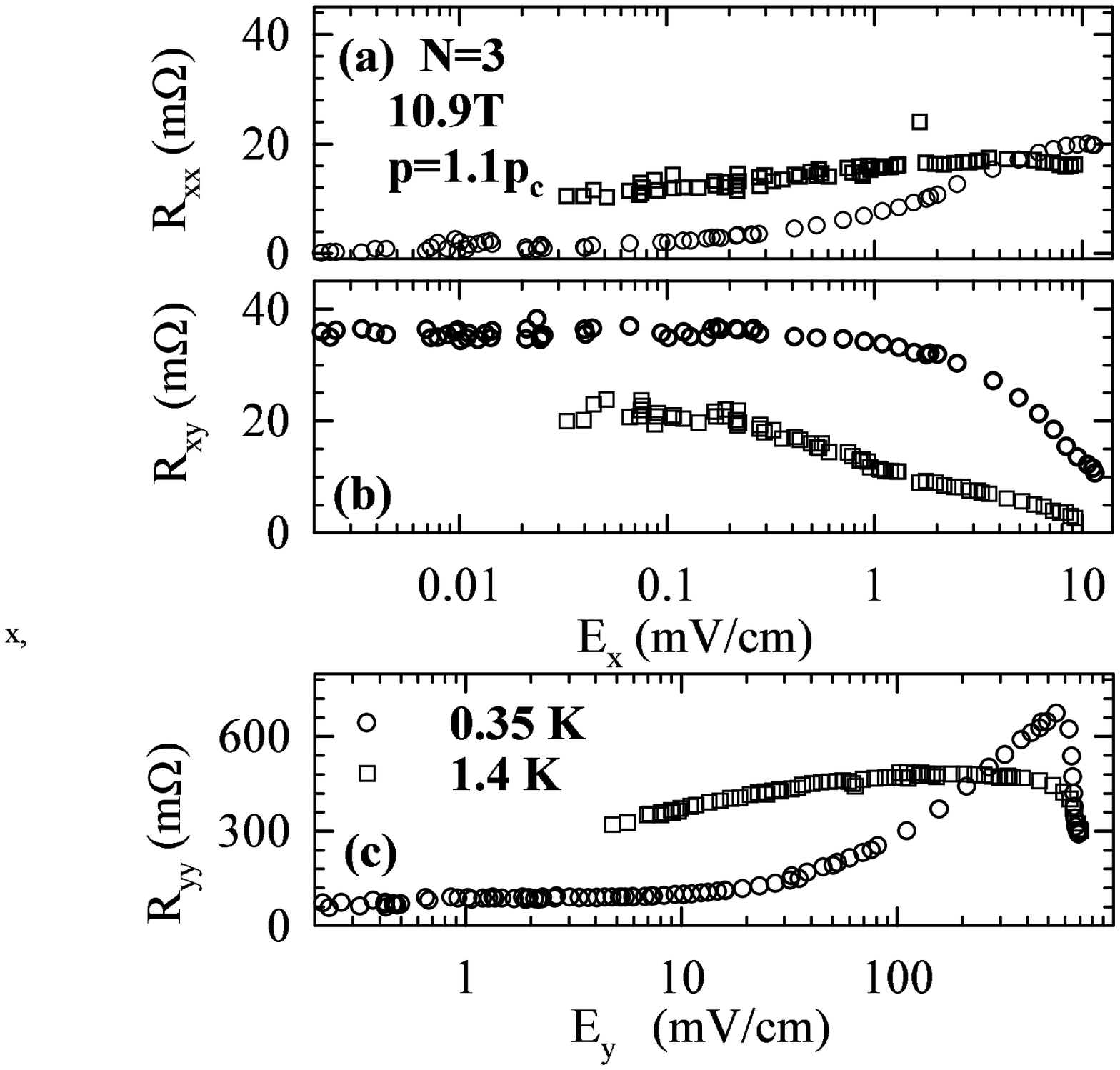}}
\caption{ Components of the resistance tensor $R_{xx}$ (a), $R_{xy}$ (b) and $R_{yy}$ (c) \vs electric field at a few selected temperatures for a high quality sample.}
\label{rhotensor}
\end{figure}

In addition, in Fig.~\ref{rhotensor} we show the electric-field dependence of three components  $R_{xx}$,  $R_{xy}$, and  $R_{yy}$ of the resistivity tensor for the high quality single crystal. We note that shown data, which correspond to the subphase N=3, were measured at pressure $p=1.1p_c$ at which measurements in the linear regime have been performed (Fig.~\ref{high},~\ref{low}). This pressure differs slightly from the one at which data displayed in Fig.~\ref{NonL9900}(a)  have been measured. Two features should be pointed out. First is that the diagonal components $R_{xx}$ and $R_{yy}$ at both temperatures 0.35 K and 1.4 K, representative of the low and the high temperature range as already described, follow the qualitatively same behaviour as a function of electric field $E_{x}$  and $E_{y}$, respectively. That is, the increase above a finite electric field and a further drop above an order of magnitude larger fields (for $R_{xx}$ see also Fig.~\ref{NonL9900}(a)). Second is that the Hall component $R_{xy}$ decreases with electric field, concomitantly with the rise of $R_{xx}$ and $R_{yy}$. It should be noted that the change of both $R_{xx}$ and $R_{xy}$ start above a finite electric field of about $(0.6\pm0.3)$~mV/cm, whereas $R_{yy}$ starts to increase above a field of about  $(30\pm10)$~mV/cm. It should be noted that $R_{yy}$  in the normal phase stayed constant up to 50 mV/cm, that is the value at least two times larger than the threshold field   (refer to threshold field values as in Fig.~\ref{rhotensor}).

Our second important result concerns the validity of chosen contact geometry  to probe the components, both linear and non-linear, of resistivity tensor in the FISDW phases. Measurements of the longitudinal resistivity $\rho_{xx}$ at low and high electric fields, that is, in the linear and non-linear regime, on low quality single crystals gave qualitatively similar results independently of the used contact configuration. Conversely, in the case of the high quality single crystals, the chosen configuration has revealed  to have a decisive influence on the experimentally observed behavior of the linear and non-linear resistivity. Precisely, results obtained in the four annular contact configuration were qualitatively and quantitatively similar to the results valid for the low quality sample like one presented in Fig.~\ref{low} and Fig.~\ref{NonL9900}(b). On the basis of this result we conclude that the four annular contact geometry is not a proper method to probe the electrical transport in the QHE state of FISDW phases. 

\section{Discussion}
We are going first to address the properties of the linear resistivity tensor, and in the second part of the paragraph we comment the behavior of the resistivity tensor in the non-linear regime.

For the high quality single crystals, the longitudinal resistivity displays, after the initial rise below    $T_{c}$, a significant decrease towards low temperatures (Fig.~\ref{high}(b)). This decrease happens concomitantly with  the gradual establishment of the well defined plateau in the Hall resistance (Fig.~\ref{high}(c)). As for the low quality samples, a decrese towards low temperatures is much less pronounced and it is not visible at all in measurements at lower magnetic fields (Fig.~\ref{low}(b)).

In what follows we apply the formalism, developed by Virosztek and Maki  (VM)~\cite{ViMa75},~\cite{ViMa90}, to deal with linear resistivity tensor for both high and low quality samples, taking into account the Quantum Hall Effect of the FISDW phases. We propose here first that QHE is readily incorporated by replacing $\sigma_{xy}$ by
\begin{equation}
\sigma_{xy}=\sigma_{xy}^{cl}+\sigma_{xy}^{Q}\label{sigclQ}
\end{equation}
where the first, classical term $\sigma_{xy}^{cl}$ has been already considered by VM, whereas the QHE term $\sigma_{xy}^{Q}$ has been obtained by Yakovenko and Goan~\cite{Yako81}. Eq.~\ref{sigclQ} is the most natural form, since $\sigma_{xy}^{Q}$ arises from the Hopf term or Chern-Simon term~\cite{Yako83}, which is neglected in the diagramatic analysis of VM,~\cite{ViMa75}.  Term $\sigma_{xy}^{Q}$ is given by
\begin{equation}
\sigma_{xy}^Q(t)=\frac{2e^2N}{h}f(t)\label{sigQ}
\end{equation}
where $f(t)$ is the dimensionless condensate density of FISDW. $f(t)$ is a function of reduced temperature $t=T/T_c$. An explicit expression for $f(t)$ depends on the order in which the limits of zero frequency $\omega\rightarrow0$ and zero momentum $q\rightarrow0$ are taken~\cite{Yako82},~\cite{ViMa90}. In the dynamic limit, where the $q\rightarrow0$ is taken first and then $\omega\rightarrow0$ is taken, the condensate density is
$$
f_d(t)=1-\int_{-\infty}^\infty{d\xi\Big(\frac{\xi}{E}\Big)^2\Big[-\frac{\partial{n_F(E)}}{\partial{E}}\Big]}
$$
where $E=\sqrt{\xi^2+\Delta^2(T)}$ is quasi-particle energy in the FISDW state, with $\xi=v_F(p-p_F)$. $v_F$, $p_F$ and $n_F$ are Fermi velocity, momentum and  distribution function, respectively. In the static limit ($\omega\rightarrow0$ first, then $q\rightarrow0$), the condensate density is
$$
f_s(t)=1-\int_{-\infty}^\infty{d\xi\Big[-\frac{\partial{n_F(E)}}{\partial{E}} \Big]}
$$
The static limit is appropriate for calculation of the magnetic field penetration depth in superconductors, which comes from the truly static Meissner effect in the thermodynamic equilibrium ({\em i.e.} $f_s$ describes superfluid density in the superconductor). It was also considered appropriate for the description of the SDW Fr\"{o}hlich conduction, while the dynamic limit was ascribed to high electric fields or high frequency conduction regimes in the SDW state~\cite{ViMa90}. On the other hand, the Hall effect is kinetic and not thermodynamic and Yakovenko {\it et al.}~\cite{Yako82} considered the dynamic limit more appropriate. They lacked experimental data to resolve this question.
Only graphical expressions for $f_s$ and $f_d$ were presented insofar (see~\cite{Yako82},~\cite{ViMa90}), and explicit analytical expressions are necessary for  proper construction  of the FISDW conductivity tensor components. Eventually, the fitting procedures define the adequate choice of the limit used.
Finally, for the fitting procedures purpose analytical expressions for $f(t)$ were obtained in the  dynamic  limit 
\begin{equation}
f_d(t)=\frac{1.3793(1-t^4)^{1/2}}{1.3793-0.3793t^4}\label{fd}
\end{equation}
and in the static limit 
\begin{equation}
f_s(t)=\frac{1-t^4}{2-t^4}\label{fs}
\end{equation}.

The diagonal components of the conductivity tensor in the SDW or FISDW phase, calculated by VM~\cite{ViMa75}, are given as 
{\setlength\arraycolsep{2pt}
\begin{eqnarray}
\sigma_{xx}(t)&=&\sigma_{nxx}n_{qp}\label{sigxx}\\
\sigma_{yy}(t)&=&\sigma_{nyy}n_{qp}\label{sigyy}
\end{eqnarray}}
where $\sigma_{nxx}$ and $\sigma_{nyy}$ are normal state (at given magnetic field) conductivity components just above $T_c$, and $n_{qp}$ is the quasi-particle density, usually given by $n_{qp}=\exp(-\Delta(T)/T))$ for systems with a gap in the quasi-particle spectrum. We replace the latter expression for  $n_{qp}$ by
\begin{equation}
n_{qp}=2(1+\exp(\Delta(T)/T))^{-1}+n_r
\label{nqp}
\end{equation}
where $\Delta(T)=\Delta(0)(1-t^4)^{1/2}$ and $\Delta(0) $ is the magnetic-field dependent single-particle gap at zero temperature inside each FISDW subphase. 

The textbook expression $n_{qp}=\exp(-\Delta(T)/T))$ is an approximation and the first term in the Eq.~\ref{nqp} is the more adequate expression of $n_{qp}$ for the SDW, in general. Explicitly, the first term is $ n_{qp}=2F(\Delta(T))$,
where $F$ is the Fermi distribution function. 
It can be derived from the $\omega\rightarrow0$ limit of the electric conductivity in SDW, since the conductivity in SDW is the same as the ultrasonic attenuation coefficient in BCS theory. In other words, the coherence factor in the conductivity in SDW is the same as the one in sound 
absorption in s-wave superconductors~\footnote{T. Tsuneto, Phys. Rev. {\bf 121}, 402, (1961).}.   The second term, $n_r$, in the Eq.~\ref{nqp} appears to be crucial for the correct description of the experimental data. It represents  the residual quasi-particle density associated with the residual conduction channel due to crystal defects or imperfections. Further, we exclude the possibility that $n_r$ is due to impurities, since impurities suppress both $T_c$ and $\Delta(0)$ rather strongly in contrast to our observation.

Components of the resistivity tensor $\rho_{ij}$ are obtained by simple inversion of the conductivity tensor
{\setlength\arraycolsep{2pt}
\begin{eqnarray}
\rho_{xx}&=&\frac{\sigma_{yy}}{\sigma_{xx}\sigma_{yy}+\sigma_{xy}^2}\label{rxx}\\
\rho_{xy}&=&\frac{\sigma_{xy}}{\sigma_{xx}\sigma_{yy}+\sigma_{xy}^2}\label{rxy}\\
\rho_{yy}&=&\frac{\sigma_{xx}}{\sigma_{xx}\sigma_{yy}+\sigma_{xy}^2}\label{ryy}
\end{eqnarray}}

Further on we assume $|{\sigma_{xy}^{cl}}/{\sigma_{xy}^{Q}}|\ll1$ for both high and low quality single crystals. Therefore,  in Eq.~\ref{sigclQ} the classical term $\sigma_{xy}^{cl}$ might be neglected. Inserting conductivity tensor components $\sigma_{ij}$, given by Eqs.~\ref{sigQ},~\ref{sigxx}, and~\ref{sigyy} (where redefined $ n_{qp}$, given by Eq.~\ref{nqp} is used) into Eqs.~\ref{rxx},~\ref{rxy}, and~\ref{ryy} we get
{\setlength\arraycolsep{2pt}
\begin{eqnarray}
\frac{\rho_{xx}(t)}{\rho_{nxx}}& = & D^{-1}=(n_{qp}+Af^2(t)n_{qp}^{-1})^{-1}\label{linRxx}\\
\frac{\rho_{xy}(t)}{\rho_{xy}(0)} & = &A f(t)n_{qp}^{-1}D^{-1}\label{linRxy}\\
\frac{\rho_{yy}(t)}{\rho_{nyy}} & = &D^{-1}\label{linRyy}
\end{eqnarray}}
where the off-diagonal parameter A is defined as
\begin{equation}
A=\frac{(\sigma_{xy}^Q(0))^2}{\sigma_{nxx}\sigma_{nyy}}
\label{A}
\end{equation}
Here $\rho_{nxx}$ and $\rho_{nyy}$ are normal state (at given magnetic field) resistivity components just above $T_c$, and $\rho_{xy}(0)$ is zero temperature Hall resistivity, effectively equal to the lowest temperature one for the same given magnetic field. When $T \ll T_c$, the denominator $D$ in Eqs.~\ref{linRxx},~\ref{linRxy}, and~\ref{linRyy} is dominated by the  $n_{qp}^{-1}$ term, and at the lowest temperatures by the $n_r$. Only at higher temperatures it comes out that the choice of the dynamic limit expression for the condensate density $f_d$, as given in the Eq.~\ref{fd},  is necessary for accurate description of the resistivity tensor behavior, {\it i.e.} for obtaining the best fits. 

\begin{figure}
\centering\resizebox{0.42\textwidth}{!}{\includegraphics*{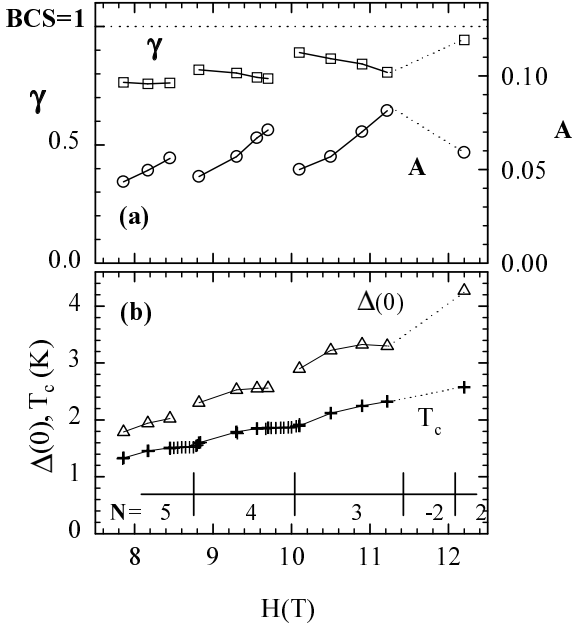}}
\caption{Magnetic field dependence of (a) off-diagonal parameter ($A$) and ratio $\frac{2\Delta(0)}{3.52T_c}$ ($\gamma$) and (b) transition temperature ($T_c$), and energy gap at 0 K ($\Delta(0)$).}
\label{Dia03}
\end{figure}

For the high quality samples fits to our experimental data for $R_{xx}$ and $R_{xy}$ shown in Fig.~\ref{high}(b),(c) are performed using Eqs.~\ref{linRxx} and~\ref{linRxy}. The free parameters are the zero temperature order parameter $\Delta(0)$, residual quasi-particle density $n_r$, and the off-diagonal parameter $A$. The fixed parameter was $T_c$, which was defined as a maximum in second derivative of resistance {\it vs.} temperature curves. $T_c$ was determined with precision of the order of  mK and $\Delta(0)$ was extracted from fits with accuracy better than 1\%. Best fits for both $R_{xx}$ and $R_{xy}$ are obtained with values of $\Delta(0)$, $A$, and $T_c$ shown in Fig.~\ref{Dia03} and  with $n_r\sim0.003$.  The fits  are in an excellent agreement with experimental data (see Fig.~\ref{high}(b),(c)). 

\begin{figure}
\centering\resizebox{0.42\textwidth}{!}{\includegraphics*{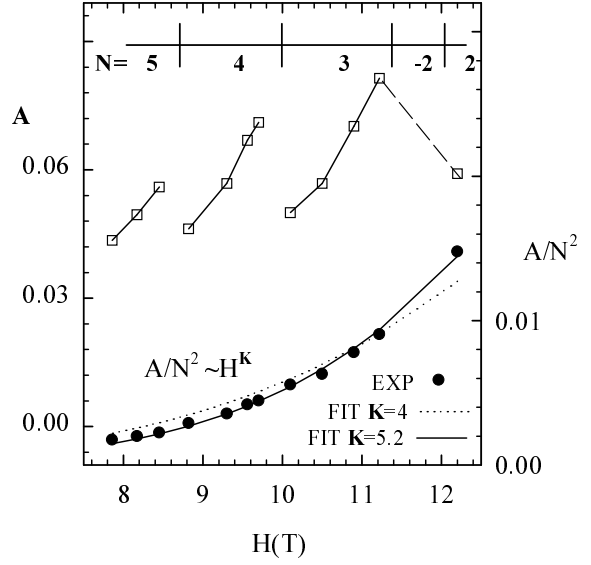}}
\caption{Magnetic field dependence of  parameter $A$,  open squares, and  parameter $A/N^2$, full circles, where $N$ is the  respective subphase index. Fits to $A/N^2 \sim H^4$, dotted line, and  to $A/N^2 \sim H^{5.2}$, solid line.}
\label{Dia04}
\end{figure}

 In principle, parameter $A$ values could be obtained directly from the experimental values of $\rho_{nxx}$, $\rho_{nyy}$, and $\rho_{xy}(0)$, for comparison with the values obtained from the fits. Still, the calculation of those three resistivity tensor components from the resistances measured is hindered by the inaccuracy in determination of the geometry of the sample and contacts. Further, it should be noted that a factor of $N^2$ may be extracted from $A$, coming from the  $\sigma_{xy}(0)$ in Eq.~\ref{A}. Therefore, in Fig.~\ref{Dia04}, we present $A/N^2$, with the bare $A$ parameter.  
 
Further on, we remind that $\rho_{nxx}$ and $\rho_{nyy}$ in Eq.~\ref{A} correspond to normal (metallic) state of  \pff~under magnetic field.  We have  observed that, {\it e.g.},  normal state magnetoresistivity $\rho_{nxx}$ is 20-40 times larger than the zero-field resistivity at the same temperature. 
In Fig.~\ref{Dia04} we show that $A/N^2$ follows a power law such as $H^K$; the best fit being obtained with $K=5.2$. Although this behavior is not that far from the textbook quadratic field dependence expected for a classical Drude metal, we recall that the magnetoresistance of Bechgaard salts and even the nature of the metallic state at low temperatures in zero magnetic field are still controversial subjects (Danner and Chaikin~\cite{Danner}, Behnia~\al{Behnia}, Balicas~\al{Balicas17}, Kriza~\al{Kriza33}).

 As  for the low quality samples, the fits  are also in an excellent agreement with experimental data (see Fig.~\ref{low}(b)). It should be noted that  fits  to Eq.~\ref{linRxx} give about the same $\Delta(0)$, and use the same fixed $T_c$ as the fits for the high quality sample.  Fitting procedure for the low quality sample gives  values  of $A$ somewhat smaller than for the high quality one. Solely one  parameter is changed significantly for the low quality sample and that is the residual quasi-particle density $n_r\sim0.10$. It should be noted that it is enhanced about 30 times compared with the value $n_r\sim0.003$ for the high quality sample. Since this parameter might be considered as a measure of the sample's  imperfections or defects level it is natural to  compare it  with the resistivity ratio $RR$, which quantifies the same feature. 
 Indeed, the ratio of $RR$ values for high and low quality samples is about 25, and is in an excellent agreement with the enhancement of $n_r$ value for low quality sample, which is around 30.
 
 In addition, in Fig.~\ref{Dia03} we show magnetic field dependence of the ratio $\gamma$
\begin{equation}
\gamma=\frac{2\Delta(0)}{3.52T_c}
\label{gamma}
\end{equation}
 We emphasize that $\gamma$ decreases continuously inside one FISDW subphase $N$ and then increases abruptly to the next higher value in subsequent phase $N-1$. A stepwise approach of $\gamma$ to the theoretical BCS value of 1 occurs. As magnetic field is increased, the BCS thermodynamics are recovered. We remind of the high experimental accuracy which provides reliable  $\gamma$ values  and allows quantitative comparison of the observed $\gamma$ dependence  with the theoretical predictions graphed by  Montambaux and Poilblanc~\cite{MoPo50}. They have based their predictions on  the standard assumption of quantized FISDW wave vector $Q_x=2k_F-NebH/\hbar c$~\cite{Yako81}.
That is, each FISDW subphase appears with a gap opened at slightly different  wave vector defined by $N$. Then, inside given subphase corresponding wave vector changes continously with the magnetic field, and this combination of steplike and continuous change is keynoted in the behavior of $\gamma$.
For the first time, these results show that the series of gaps opening in the band structure has to be taken into account in the termodynamics of the FISDW states as predicted by different authors~\cite{MoPo50,Yama86,ViChMa,PHML}. 

We would like to point out that the perfect correspondence between experimental and theoretically predicted behavior of $R_{ii}$ \vs $T$ crucially depends on two factors: the choice of dynamic limit for the condensate density (Eq.~\ref{fd}) and  proper formulation of quasi-particle density (Eq.~\ref{nqp}). It also proves that the classical term $\sigma_{xy}^{cl}$ is negligible in the FISDW subphases, independently of the quality of the nominally pure samples.  We would also like to mention that Yakovenko {\it et al.} were not sure which condensate density; the dynamic limit or the static limit,  should be used, though they speculated that the former should be a proper choice~\cite{Yako82}. Indeed, the present experiment shows clearly that the dynamic limit is the correct choice.

Further, the success in accomodating both high and low quality samples within one simple formulation of magnetic field dependent linear resistivity tensor for  each FISDW subphase enables us to conclude on the development of quantized Hall effect in both high and low quality samples.  
Any level of crystal imperfections or defects  activates   conduction channels via localized states inside the gap. Residual density of quasi-particles in the sample prohibits the complete disappearance of diagonal conductivity components $\sigma_{xx}$ and $\sigma_{yy}$ as temperature approaches to zero. Consequently, since  the diagonal resistivity component $\rho_{xx}$ is given by Eq.~\ref{rxx}, we expect a measurable  residual ({\it i.e.} at $T=0$~K) $\rho_{xx}$ value. This value is governed solely by the  residual carrier density. It should be noted that this result  holds at best if the QHE state is considered to be preserved  in all nominally pure samples, that is if the Hall conductivity $\sigma_{xy}$ attains the theoretically expected large quantized value at lowest temperatures in the low quality, as well as in the high quality samples. Therefore, although displaying apparently different behaviors of  $R_{xx}$ \vs $T$ both high and low quality samples should be regarded at equal footing concerning the development of QHE. In addition, the experimental result that the annular contact configuration revealed not to be a proper method to probe the resistivity tensor in the QHE state has the similar origin. That is, annular contacts make the short circuit in the $b'$ direction, so that $\sigma_{yy}$ increases substantially, and the residual $\rho_{xx}$ value increases again. 

Now, we address the resistivity  tensor behavior in the non-linear regime. 
First, we show that the observed nonlinearities can not be attributed to hot electron effects or to the destruction of the FISDW condensed state by electric field. Non-linear behaviour for $\rho _{xx}$ and $\rho _{xy}$, and for $\rho _{yy}$  is switched on above finite fields not larger than 1 mV/cm and 30 mV/cm, respectively. We note that even an electric field of 30mV/cm provides much smaller energy ($\epsilon$) on a microscopic length scale than $k_BT$ and $\Delta(0)$, namely, $\epsilon/k_BT<0.04$ and $\epsilon/\Delta(0)<0.01$. This can be estimated as follows. The electronic mean free path of the (TMTSF)$_2$X salts is of the order of 1 nm at RT. From the resistivity ratio for the high quality sample of about 1000 we inferred the upper limit for the mean free path of 1000 nm, at zero magnetic field. Thus the maximum energy obtainable from electric field  is about 0.01K. 

Furthermore, our results for the low quality single crystals are reminiscent of those in spin density wave systems where the nonlinearities have been attributed to the sliding SDW becoming depinned in high enough electric fields, whereas for the high quality single crystals one important difference has been found. That is, in the latter case the decrease of the longitudinal resistivity $\rho _{xx}$ has been found only after an initial rise above the threshold field $E_T$. It should be noted that $\rho_{yy}$ shows the qualitatively same behaviour, as previously reported by Balicas. 

For each FISDW subphase there also should exist a translational spin-density wave mode, which is theoretically expected to give a collective conductivity similar to a regular SDW case, when the QHE is neglected (VM~\cite{ViMa75}). Indeed, a decrease of the longitudinal resistivity above a finite threshold field has been observed in the low quality single crystals in which the QHE state was masked by a weak disorder. The QHE nature of FISDW has been taken into account by Yakovenko and Goan~\cite{Yako81}.  They treated the FISDW motion in terms of the single oscillator model where both damping and pinning frequency are introduced phenomenologically.
 Their most important result is that the FISDW motion in a non-stationary regime gives rise to the longitudinal conductivity that exactly cancels the Hall conductivity. Further, Yakovenko did not expect to observe such an effect in DC electric fields, since in the latter case the acceleration of the FISDW rapidly vanishes and a steady flow is stabilized. The following experimental observations indicate the validity of this theoretical prediction also in DC electric fields. First, we recall that we have observed a decrease of the Hall resistivity $\rho _{xy}$  that is switched on above the threshold field concomitantly with an increase of the longitudinal resistivity $\rho _{xx}$. Second, the magnitude of $\rho _{xx}$ and $\rho _{xy}$ changes as a function of electric field are strongly sensitive to a level of the QHE order. That is, at higher temperatures above the peak in the $R$ \vs $T$ low field curves  where the QHE state is not yet fully developed, $ \rho _{xx}$ and $\rho _{xy}$ showed much less pronounced increase and decrease at twice a threshold field, respectively. The increase of $\rho _{xx}$  should be attributed to the sliding FISDW since Shapiro interference, immediately above the  threshold electric field, was found~\cite {Balicas08}. At high enough electric fields the QHE regime is destroyed and an usual decrease of  $\rho _{xx}$ corresponding to a longitudinal conductivity rise should be expected in accordance with our observations.

Finally, as far as the temperature dependence of the threshold field for the longitudinal resistivity is concerned, we have found that $E_T$ decreases with increasing temperature, independently of the sample quality. In addition, it should be noted that the low quality samples showed an order of magnitude larger $E_T$ for $\rho_{xx}$ than the high quality one. Here we would like to point out that due to finite nondiagonal components of the resistivity tensor we cannot use the theoretical model for a regular SDW~\cite{ViMa90},~\cite{Tomic68} to compare our results with the theoretical predictions in order to deduce the dominant pinning mechanism.

In conclusion, new data for the electric field dependent transport and Hall effect in the FISDW phases of \pff~have been obtained. The conductivity at low fields, as well as the sliding conductivity above a finite threshold electric field have been understood in terms of a microscopic model which takes into account explicitly the QHE nature of these FISDW phases. Such a model provides a reasonable interpretation for mutually inconsistent data obtained earlier by several authors. A careful experimental study of the conductivity activation gap in each FISDW sub-phase reveals that the thermodynamics is governed by a magnetic field dependent nesting vector as predicted theoretically~\cite{MoPo50,Yama86,ViChMa,PHML}.
\end{document}